\documentclass[aps,prl,twocolumn,showpacs,superscriptaddress,groupedaddress]{revtex4}  
\usepackage{graphicx}  
\usepackage{dcolumn}   
\usepackage{bm}        

\usepackage{amsfonts,epsfig,amsmath,amsthm,amssymb,graphics,verbatim}

\def\R{\mathbb{R}}

\def\cK{\mathcal{K}}

\def\cX{\mathcal{X}}

\def\ra{\rightarrow}
\def\I{\infty}

\newcommand{\be}{\begin{equation}}
\newcommand{\ee}{\end{equation}}
\newcommand{\benn}{\begin{equation*}}
\newcommand{\eenn}{\end{equation*}}
\newcommand{\bea}{\begin{eqnarray}}
\newcommand{\eea}{\end{eqnarray}}
\newcommand{\beann}{\begin{eqnarray*}}
\newcommand{\eeann}{\end{eqnarray*}}

\begin{document}

\title{Time-Scale and Noise Optimality in Self-Organized Critical Adaptive Networks}
\date{\today}
\author{Christian Kuehn}
\affiliation{Institute for Analysis and Scientific Computing, Vienna University of Technology, Wiedner Hauptstrasse 8-10, 1040 Vienna, Austria}
\affiliation{Max Planck Institute for the Physics of Complex Systems, Noethnitzer Strasse 38, 01187 Dresden, Germany}
\begin{abstract}
Recent studies have shown that adaptive networks driven by simple local rules can organize into ``critical'' global steady states, providing another framework for self-organized criticality (SOC). We focus on the important convergence to criticality and show that noise and time-scale optimality are reached at finite values. This is in sharp contrast to the previously believed optimal zero noise and infinite time scale separation case. Furthermore, we discover a noise induced phase transition for the breakdown of SOC. We also investigate each of the three new effects separately by developing models. These models reveal three generically low-dimensional dynamical behaviors: time-scale resonance (TR), a new simplified version of stochastic resonance - which we call steady state stochastic resonance (SSR) - as well as noise-induced phase transitions.    
\end{abstract}

\maketitle

The concept of self-organized criticality (SOC) was first proposed by Bak, Tang and Wiesenfeld \cite{BakTangWiesenfeld} using the abelian sandpile model. The idea is that a dynamical system organizes without external influence to a special state near a continuous phase transition \cite{Jensen} i.e. to a state of marginal stability \cite{Turcotte} which could provide optimal information processing \cite{HaldemanBeggs}. A precise definition of SOC does not seem to exist \cite{Turcotte}, but all models for SOC exhibit dynamics on different time-scales \cite{Jensen}. Recent studies have shown that SOC can be observed in adaptive networks (see e.g. \cite{BornholdtRohlf,ChristensenDonangeloKoillerSneppen,PaczuskiBasslerCorral}) where dynamical rules influence the network topology and vice versa. The focus for SOC in the adaptive networks has been to analyze the critical point and applications to neuroscience \cite{BornholdtRoehl,LevinaHerrmannGeisel}. Here we take a different viewpoint and focus on the convergence towards SOC, which is very important as a return from nearby states to SOC is required to actually \emph{use} the criticality of the state. We are interested in optimality of the passage to SOC which is particularly interesting the context of forming optimal networks in the brain \cite{Bassettetal}. As a case study we consider an SOC network proposed by Bornholdt and Rohlf (BR) \cite{BornholdtRohlf}. We expect the phenomena in this model to apply to a wide variety of adaptive network SOC models; however, this claim still needs to be justified with future work. Effects found in full network simulations are going to be studied by developing simple models. This approach of isolating generic effects via simple low-dimensional mathematical models has a long history and has been particularly successful in the context of dynamical systems \cite{GH}. For SOC adaptive networks this approach has not yet been utilized as much as we believe is necessary. A mathematically rigorous derivation for SOC adaptive networks in a mean-field limit does not seem to exist yet; hence we must take the approach via numerics and modelling. We define a slightly modified version of the adaptive network in \cite{BornholdtRohlf}. Consider a directed network with nodes $\{v_i(t)\}_{i=1}^N\in\{\pm 1\}$ and edges $\{e_{ij}(t)\}_{i,j=1}^N\in\{-1,0,+1\}$ without loops. The initial network $G(t=0)=:G(0)$ at $t=0$ is chosen as a random graph with random node values. Consider the connection matrix $E(t)=\{e_{ij}(t)\}_{i,j=1}^N$ and define 
\be
f(t)=E(t)v(t)+\mu v(t)+\sigma r_i 
\ee
where $r_i\sim \mathcal{N}(0,1)$ represents noise with $\sigma\geq 0$ and $\mu\in[0,1]$ controls the influence of the current state of the node on its new state. The case $\sigma=0=\mu$ corresponds to \cite{BornholdtRohlf}. Then the node dynamics is given, via parallel update, by 
\be
\label{eq:BR_node1}
v_i(t+1)=\left\{\begin{array}{ll}
\text{sgn}[f_i(t)] & \text{ if $f_i(t)\neq 0$,}\\
v_i(t) & \text{ if $f_i(t)= 0$.}\\
\end{array}
\right.
\ee  
Possible applications for this setup are neural networks \cite{MeiselGross}, opinion formation \cite{HolleyLiggett} and collective behaviour \cite{HuepeZschalerDoGross} where the rule \eqref{eq:BR_node1} corresponds to nearest-neighbours information transmission.
 
\begin{figure}[tbp]
	\centering
		\includegraphics{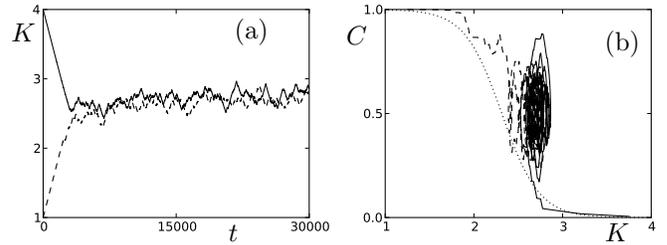}
		\caption{\label{fig:fig1}Simulation of the BR model with initial conditions $K_{ini}=1$ (dashed curve) and $K_{ini}=4$ (solid curve) for $(N,\epsilon,\delta,\mu,\sigma)=(1000,10^{-3},10^{-6},10^{-6},0)$. (a) Time series of the average connectivity $K$. (b) Moving average of the curves from (a) in $(K,C)$-space. The dotted curve indicates the phase transition obtained in \cite{BornholdtRohlf} (for $N=1024$).}
\end{figure}
   
After $T_v$ node dynamics steps \eqref{eq:BR_node1}, the average activity of each node over the last $T_a:=\lfloor T_{v}/2\rfloor$ updates is measured
\be
\label{eq:BR_act}
A_i:=\frac{1}{T_{v}-T_a}\left[\sum_{t=T_a}^{T_{v}}v_i(t)\right].
\ee
Nodes with $|A_i|=1$ indicate frozen states, unchanged for a long time, while $|A_i|<1$ are active nodes. Denote the fraction of frozen nodes by $C$. 
Choose a site $i$ at random and calculate $A_i$ using \eqref{eq:BR_act}. Let $0\leq\delta\leq 1$ be a parameter; if $|A_i|>1-\delta$ then $i$ receives a new edge $e_{ij}$, with randomly chosen edge weight in $\{\pm 1\}$, from a node $j$ chosen at random. If $|A(i)|\leq 1-\delta$ then one of the existing nonzero edges is deleted. In the context of neural networks frozen states are activated via new connections while active nodes can be reduced in activity; see \cite{MeiselGross} for an interpretation in terms of synaptic plasticity. After the topological update the dynamics switches again to $T_v$ node dynamics steps \eqref{eq:BR_node1}. It is natural \cite{GrossBlasius} to assume a time scale separation between the topological (slow) and the node (fast) dynamics which requires $T_v\gg1$; we define $0<\epsilon:=1/T_v\ll1$. For example, individual neuron dynamics is faster than synaptic plasticity. Bornholdt and Rohlf observed that random graphs with different initial connectivity $K_{ini}$ converge to a network with the same connectivity $K_{c}(N)$ (see Figure \ref{fig:fig1}(a)) where $K_c(N)\ra 2$ as $N\ra \I$. For fixed connectivity $K$ and fixed number of nodes $N$, they averaged over different static topologies revealing a phase transition of $C(K,N)$ from $C\approx 0$ to $C\approx 1$ near $K_c$ upon decreasing $K$; see Figure \ref{fig:fig1}(b). Hence the adaptive network self-organizes into a critical state $(K_c,C_c)$ using the combination of topological and node dynamics. 

\begin{figure}[htbp]
	\centering
		\includegraphics{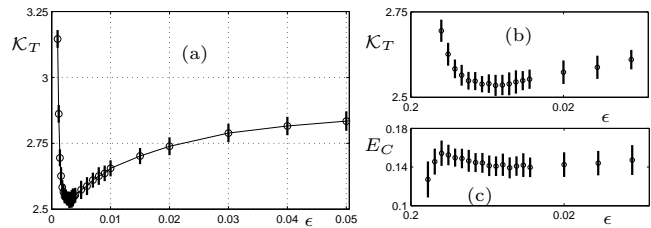}
		\caption{\label{fig:fig2}Full network simulation (dots), averaged over $100$ samples $G(0)$ with one standard deviation bars for $(N,T,K_{ini},\delta,\mu,\sigma)=(1000,6\cdot 10^4,4,10^{-6},10^{-6},0)$. (a) Target connectivity $\cK_T=\cK_T(\epsilon)$; interpolation of the data is shown. (b) Zoom near the global minimum from (a). (c) Frozen component error $E_C=E_C(\epsilon)$.}
\end{figure}

We are interested in the dependence of SOC on the time scale separation and noise where the key object of study are finite-time trajectories of moving-average macroscopic variables $(K(t),C(t))$ for $t\in[0,T]$. Assume that $K_{ini}$ for $G(0)$ is given with $K_{ini}\not\approx K_c$ so that the network has to self-organize. First, we investigate the effect of $0<\epsilon\ll1$ for $T=60000$. Define $(K_T,C_T):=(\langle K(t)\rangle_{[T/2,T]},\langle C(t)\rangle_{[T/2,T]})$ where $\langle\cdot\rangle_{[T/2,T]}$ denotes the time average for $t\in[T/2,T]$ i.e. we view the trajectory in $[T/2,T]$ as being near its target state. Note that $K_T,C_t$ are random variables depending on the initial sample $G(0)$. Figure \ref{fig:fig2} shows $K_T$ for an average over 100 samples of $G(0)$; averaging over $G(0)$ is indicated by calligraphic letters e.g. $\langle K_T\rangle_G=\cK_T$. Define the errors
\be
E_{X}:=\frac{2}{T}\sum_{t=0}^{T/2} |X(t)-\cX_T|,\qquad \text{for $X\in\{K,C\}$.}
\ee

Figure \ref{fig:fig2}(a)-(b) displays a global minimum $\cK_T(\epsilon_{min})$ upon time scale variation; this is an effect caused by the high connectivity initial condition $K_{ini}=4$ and the curve is expected to be monotone increasing for $K_{ini}$. There is always a unique finite $\epsilon$ optimal for the criterion $\min_\epsilon |\cK_T(\epsilon)-K_c|$ of trajectories being close to the critical connectivity. The error $E_C(\epsilon)$ in Figure \ref{fig:fig2}(c) also seems to depend non-monotonously on $\epsilon$ and is always expected to yield a unique finite $\epsilon$ optimality if we start away from the critical point. These observations lead to the main conclusion that SOC systems have \emph{optimal operational time scales}.

As stated in the introduction, after identifying an effect, we aim to develop a simple abstract representation of the dynamical phenomenon. As a model problem consider the fast-slow system
\be
\label{eq:genfs}
\begin{array}{lclcr}
\frac{dx}{dt}&=&x'&=&f(x,y),\\
\frac{dy}{dt}&=&y'&=&\epsilon g(x,y).
\end{array}
\ee 
For illustration purposes, we introduce one of the simplest systems with time scale optimality given by
\be
\label{eq:fs1}
f(x,y)=(y-1)^2-x,\quad g(x,y)=1-y.
\ee

\begin{figure}[htbp]
	\centering
		\includegraphics{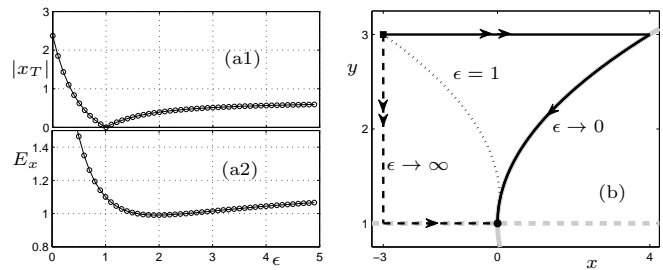}
		\caption{\label{fig:fig3}Illustration of time scale optimality for model problem \eqref{eq:genfs}-\eqref{eq:fs1}. (a) $|x_T|$ and error $E_x(\epsilon)$ for trajectories starting at $(x(0),y(0))=(-3,3)$ (square) with integration time $t\in[0,2]=[0,T]$. (b) Geometry in phase space. Critical manifolds as $\epsilon\ra 0$ (dashed grey), $\epsilon\ra \I$ (solid grey) with associated singular trajectories (black dashed/solid) are shown. The dotted curve is numerical integration for $\epsilon=1$.}
\end{figure}

This system has a globally asymptotically stable steady state $(0,1)=:(x_c,y_c)$ which we think of as the SOC state. Figure \ref{fig:fig3}(a1) shows $|x_T|$, calculated for $T=2$ as before, which exhibits optimal criticality near $\epsilon\approx 1$ when $x_T\approx x_c$. Figure \ref{fig:fig3}(a2) shows the error $E_x(\epsilon)=E_x$ displaying a clear global minimum. Time scale optimality is a phenomenon that has not been considered so far in the geometric fast-slow systems theory \cite{Jones}. However, there is an easy geometric explanation illustrated in Figure \ref{fig:fig3}(b). If $\epsilon\ra 0$ then $x$ is fast and $y$ is slow. The critical manifold $S_f=\{f=0\}$ is normally hyperbolic attracting with respect to $x$ as $D_xf|_{S_f}=-1<0$. Slow flow trajectories on $S_f$ converge to $(x_c,y_c)$. If $\epsilon\ra \I$, the situation is reversed with $x$ slow and $y$ fast. The critical manifold $S_g=\{g=0\}$ is normally hyperbolic attracting as $D_yg|_{S_g}=-1<0$. After a fast initial segment trajectories on $S_g$ converge to $(x_c,y_c)$. Figure \ref{fig:fig3}(b) also shows a numerically integrated trajectory for $\epsilon=1$. Fenichel theory \cite{Fenichel4} implies that for $0<\epsilon\ll1$ or $1\ll1/\epsilon$ the geometric description persists and $S_{f,g}$ perturb to invariant manifolds $S_{f,g}^\epsilon$. In a neighbourhood of $S_{f,g}^\epsilon$ dynamics is slow which explains the shape of the functions $|x_T(\epsilon)|$ and $E_x(\epsilon)$.

Obviously different slow manifold geometries can produce various optimality curves. Furthermore, the simple description of the time scale optimality phenomenon suggests that it may have a much wider relevance for multi-scale systems that occur frequently in various applications. Therefore we suggest a new name for the effect: \emph{time-scale resonance} (TR). 

\begin{figure}[htbp]
	\centering
		\includegraphics{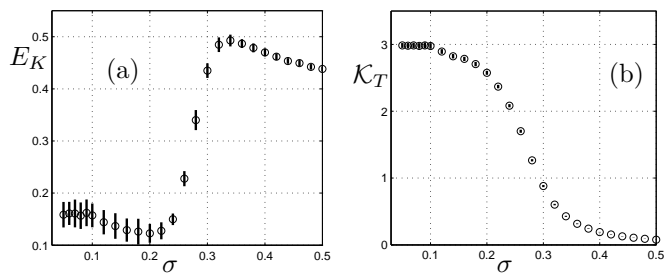}
		\caption{\label{fig:fig4}Full network simulation (dots), averaged over $100$ samples $G(0)$ with one standard deviation bars for $(N,T,K_{ini},\delta,\mu,\epsilon)=(1000,10^5,4,0.5,0.5,0)$. (a) Connectivity error $E_K=E_K(\sigma)$. (b) Phase transition in $\cK_T(\sigma)$ upon noise variation.}
\end{figure}

It was observed in \cite{BornholdtRohlf} that the adaptive SOC network is ``robust'' against small noise. Here we take this idea one step further and consider a much broader range of noise strengths. Figure \ref{fig:fig4}(a) shows the error $E_K(\sigma)$ where we first focus on $\sigma\leq0.25$. In the range $\sigma\in[0,0.25]$ we observe a noise-optimal minimum near $\sigma\approx 0.2$. This suggests the possibility of \emph{optimal noise} for SOC. It is well-known that noise can play a constructive role via effects such as stochastic resonance (SR) \cite{BenziSuteraVulpiani,NicolisNicolis} which has various applications \cite{GammaitoniHaenggiJungMarchesoni}. In SR noise-optimality is usually defined for coherence statistics along periodic orbits \cite{Lindneretal} which is different from the SOC situation where convergence to a critical state is observed. We shall explain with two models easy possibilities of noise-optimality for convergence to a steady state. First, consider \eqref{eq:genfs} with
\be
\label{eq:delayed}
f(x,y)=yx-x^3+\tilde{\sigma}\xi(t),\quad g(x,y)=(x^*-|x|)
\ee 

\begin{figure}[htbp]
	\centering
		\includegraphics{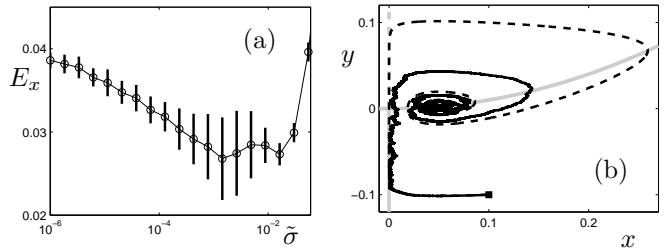}
		\caption{\label{fig:fig5}Numerical integration for \eqref{eq:delayed} with $(\epsilon,x^*,T)=(0.01,0.05,2000)$ and initial condition $(x(0),y(0))=(0.1,-0.1)$. (a) Error for ``self-organization'' depending on the noise intensity. (b) Geometry in phase space: The critical manifold $\{f=0\}$ (grey) and two trajectories for $\sigma=0$ (dashed, black) and $\sigma=10^{-3}$ (solid, black) are shown starting at $(x(0),y(0))$ (black square).}
\end{figure}

where $\langle\xi(t)\xi(t')\rangle=\delta(t-t')$ is white noise and $0<\tilde{\sigma},x^*\ll 1$ are parameters. The SDE defined by \eqref{eq:delayed} is invariant under the map $x\mapsto -x$ so we consider it only on the quotient space $(x,y)\in\R^2/(x\mapsto-x)=\R^+_0\times \R$. Observe that $yx-x^3$ is the normal form vector field for a pitchfork bifurcation with parameter $y$. In contrast to the classical delayed pitchfork bifurcation, there exists a steady state $(x_c,y_c)=(x^*,(x^*)^2)$ which represents the SOC state lying near the bifurcation point. Figure \ref{fig:fig5}(a) shows $E_x(\tilde{\sigma})$ computed from numerical integration which has a global minimum at finite noise. The geometric understanding is based on shortening of bifurcation delay due to noise \cite{BerglundGentz6} which is well understood; see Figure \ref{fig:fig5}(b). In our case, the additional steady state $(x_c,y_c)$ near the bifurcation is the novel feature that shows how shortening of bifurcation delay can lead to noise-optimality. It is very important to note that the mechanism is \emph{local} since we could let $x^*\ra 0^+$ and observe noise optimality for trajectories in an arbitrarily small neighbourhood of $(x,y)=(0,0)$. We refer to the general concept of noise-optimality for convergence as \emph{steady-state stochastic resonance} (SSR). 

There are many other possible geometries to obtain SSR. As a second example, consider the excitable systems case \cite{FitzHugh} with the vector fields in \eqref{eq:genfs} given by
\benn
f(x,y)=-\frac23y-\frac{x^3}{3}+x, \qquad g(x,y)=x-y+1+\sigma\xi(t)
\eenn
with initial conditions $(x(0),y(0))=(-2,-1-\omega)$ for some small $\omega>0$. Then it is easy to deduce from the \emph{global} geometry that SSR occurs as small noise in the slow variable can avoid the large excursion/spike towards $x>0$ and make sample paths converge directly to the globally asymptotically stable equilibrium of the deterministic system. For the same initial conditions and $\sigma=0$, we can also get TR which yields an interesting two-parameter interplay which will be considered in future work.     

We return to the full network simulation for SOC shown in Figure \ref{fig:fig4}. In Figure \ref{fig:fig4}(b) a \emph{noise-induced phase transition} can be observed between a final SOC state with networks $K_T\approx 3$ for low noise and $K_T\approx 0$ for high noise with a sharp transition near $\sigma\approx 0.25$. The effect of noise-induced phase transitions - and in particular the failure to reach the classical SOC state (here $K_c=2$) - has not been previously observed in SOC adaptive networks. The basic explanation is that higher noise values make the sequence of observed states $v_i(t)$ for $t\in[T_a,T_v]$ more random. This implies that a node $i$ is viewed as ``active'' and $|A_i|<1-\delta$ is more likely to hold which results in edge deletion. Therefore SOC is not robust against arbitrary noise $\sigma>0$. It persists up to a critical value $\sigma=\sigma_c$ which is the expected behaviour e.g. for any biological systems displaying SOC \cite{MeiselGross}.

\begin{figure}[htbp]
	\centering
		\includegraphics{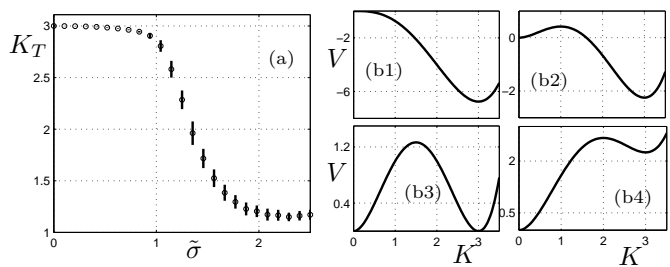}
		\caption{\label{fig:fig6}Numerical integration for \eqref{eq:noise} with $T=2000$ and initial condition $K(0)=3$. (a) Error for ``self-organization'' depending on $\tilde{\sigma}$. (b1)-(b4) Potentials $V=V(K)$ for different $\tilde{\sigma}\in\{0,1,1.5,2\}$.}
\end{figure}

To model the noise-induced phase transition, consider the connectivity $K(t)$ as evolving in an energy landscape in continuous time
\be
\label{eq:noise}
\frac{dK}{dt}=-\nabla V(K;\tilde{\sigma})+\tilde{\sigma}\xi(t),\qquad K\geq 0.
\ee 
The vector field \eqref{eq:noise} can be thought of as representing the rates of change in the infinite network limit $N\ra \I$. The adaptivity of the network suggests that local noise may also change the global potential $V$ which we choose as
\benn
V(K;\tilde{\sigma}):=\frac{K^4}{4} - \frac{K^3}{3} (3 +\tilde{\sigma}) +\frac32 K^2 \tilde{\sigma}
\eenn
as shown in Figure \ref{fig:fig6}(b). For $\tilde{\sigma}=0$ the ODE \eqref{eq:noise} has a stable steady state at $K_c=3$ for $K>0$; we choose the initial condition always as $K(0)=3$. It is important to note that the system \eqref{eq:noise} without the white noise term will always stay at the stable steady state $K_c$ for $\tilde{\sigma}\in[0,3)$.  

In Figure \ref{fig:fig6}(a) a noise-induced phase transition for $K_T$ is observed for $\tilde{\sigma}\approx 1.5$. Obviously there are many other models - discrete and/or continuous time - that may produce a similar effect. Here we have just provided one of the simplest model examples. 

In summary, we have found interesting time scale and noise optimality in convergence to SOC as well as a noise-induced phase transitions for full network simulation of the BR model. We described each of the three effects by simple abstract models hinting at interesting generic phenomena such as time scale resonance (TR), steady state stochastic resonance (SSR) as well as the possibility of noise-induced phase transitions in varying potentials. If our results turn out to hold for adaptive SOC models in applications we can draw several conclusions: (a) for processes near criticality one must determine the finite optimal time scales of these processes relative to each other, (b) information processing in complex systems near steady-state criticality can exhibit noise optimality providing another fundamental mechanism besides SR and (c) evolution may potentially drive a system towards finite optimal noise and time scale values. The last conclusion is a potential hint why it is currently problematic to match neural network models to experiment as the singular limt $(\epsilon,\sigma)\ra(0,0)$ is often the analytical starting point. 

\textbf{Acknowledgements:} I would like to thank Pawel Romanczuk and two anonymous referees for helpful comments that lead to improvements of the paper.
 
\bibliography{../my_refs}

\end{document}